# Three Dimensional Gravity and M-theory

Michael McGuigan

Brookhaven National Laboratory
Upton, NY 11973
mcguigan@bnl.gov

Abstract

It is well known that string theory can be formulated as two dimensional gravity coupled to matter. In the 2d gravity formulation the central charge of the matter together with a hidden dimension from the conformal factor or Liouville mode determines the Target space dimension. Also the vacuum amplitude of the 2d gravity formulation implies important constraints on the Target space theory associated with modular invariance. In this paper we study a three dimensional gravity approach to M-theory. We find that there are three hidden Liouville type fields coming from the 3d gravity sector and that these together with the number of zero modes of the matter fields determine an eleven dimensional Target space of M theory. We investigate the perturbative vacuum amplitude for the 3d gravity approach to M theory and constraints imposed from SL(3,Z) modular invariance using a method of Dolan and Nappi together with a sum over spin structures which generalizes the SL(2,Z) invariance found in string theory. To introduce gauge fields in M-theory we study the vacuum amplitude on a three annulus and introduce interactions with two dimensional matter on a boundary in analogy with the introduction of gauge fields for open string theory. We study a three dimensional version of M-theory from the 3d gravity perspective and show how it relates to two dimensional type 0A string theory described by a 2d superLiouville theory with c=1 matter and, on manifolds with boundary, to a E8xSO(8) 2d heterotic string. We discuss a nonperturbative 3d gravity approach to M-theory and the expansion about $(e,\omega) = 0$ in the Chern-Simons gauge formulation of the theory. Finally we study the interaction of fermionic matter with 3d gravity in order to investigate the origins of conformal dimension and Liouville effective action from the 3d gravity perspective.

# I. Introduction

The 2d gravity formulation of string theory remains a powerful approach to the subject beginning with Poyakov's work [1] and yielding nonperturbative results at least for two Target space dimensions. The basic ideas of the approach include the introduction of a hidden Liouville field remaining after fixing the conformal gauge, the coupling of 2d gravity to matter where the matter+Liouville describe a Target spacetime, the relation between 2d gravity topology change and target space scattering, the relation between 2d gravity SL(2,Z) modular invariance and absence of gauge and gravitational anomalies, and a consistency condition relating the conformal anomaly of matter with the variation of a Liouville action in the conformal gauge [2,3].

M-theory is an eleven dimensional theory whose compactification on a circle reduces to ten dimensional IIA superstring theory with coupling constant $g_s$ related to the radius of the compact dimension [4]. As M-theory reduces to string theory in a certain limit, it is a more general theory than string theory and can be used to relate different string theories which appear distinct after compactification from eleven dimensions. In this paper we investigate a 3d gravity approach to M-theory. We shall attempt to generalize the approach of 2d gravity to string theory to M-theory by introducing three hidden Liouville fields of 3d gravity remaining after gauge fixing, the coupling of matter to 3d gravity to describe a target spacetime, the notion of vacuum amplitude SL(3,Z) modular invariance and sum over spin structures to describe and constrain the structure of the theory.

Already in string theory we have indications that 3d theories play an important role. In [5] Witten and [6] Moore and Seiberg introduced the concept of a thickened world-sheet and showed that the Hilbert space of a 3d Chern-Simons gauge theory is identified with conformal blocks of a WZNW used in compactifications of string theory. In [7] Verlinde demonstrated that physical states of 3d SL(2,R) Chern-Simons generate Virasoro conformal blocks of string theory in the 2d gravity formulation. Finally in [8] Carlip and Kogan and Ferriera and Kogan use Topologically massive gauge theory and gravity in 3d with a conformal theory on a 2d boundary to construct a topological membrane approach to string theory.

Also three dimensional gravity possesses powerful nonperturbative methods like the Chern-Simons gauge gravity [9,10], Lattice 3d gravity [11], phase space path integral [12], sum over topologies [13] and even Matrix models [14]. If a 3d gravity approach to M-theory is valid one can hope to use these methods to attack many important physical questions surrounding the theory in the strong coupling regime such as those involving symmetry breaking.

This paper is organized as follows. In section I we give a basic introduction. In section II we show that there are three Liouville type fields coming from the 3d gravity sector and that these together with the number of zero modes of the matter fields determine an Target space of M-theory. In section III we investigate the vacuum amplitude for the 3d gravity approach to M-theory theory and constraints imposed from SL(3,Z) modular invariance which generalizes the SL(2,Z) invariance found in string theory. Summing over spin structures we generalize the NSR sectors of 2d supergravity coupled to matter. In section IV we introduce gauge fields in Target space in the 3d gravity approach to M-theory through interactions with two dimensional matter on a boundary in analogy with the introduction of gauge fields for open string theory. In section V we discuss a three dimensional Target space M-theory and show how it related pure three dimensional gravity and supergravity without matter. In section VI we discuss the expansion about e=0 for 3d gravity with and without matter and how this can be applied to a nonperturbative approach to the vacuum amplitudes of M-theory. In section VII we discuss insertion of two stress tensors in a fermion loop in order to investigate the emergence of conformal dimension and 2d Liouville gravity from 3d gravity plus matter. Finally in section VIII we discuss our main conclusions of the paper.

II Three Liouville fields of three dimensional gravity

In 2d gravity formualtion of string theory the Liouville field can be obtained most easily in the conformal gauge. There the 2d metric takes the form:

$$g_{\mu\nu} = \begin{pmatrix} e^\phi & 0 \\ 0 & e^\phi \end{pmatrix}$$

(2.1)

where $\phi$ is the Liouville field. The number of degrees of freedom of the theory are $3-2-2=-1$ where the negative contributions are from two gauge conditions $g_{11} = g_{22}, g_{12} = 0$ and two ghost fields.

In 4d gravity E. and H.Verlinde [15] have used a similar a gauge to expose the two gravitational degrees of freedom of the 4d theory. In that case the 4d metric takes the form:

$$g_{\mu\nu} = \begin{pmatrix} h & 0 \\ 0 & h' \end{pmatrix}$$

(2.2)

where $h$ and $h'$ are two index metrics. For 4d gravity the degrees of freedom are $10-4-4=2$ where the negative contributions are from the four gauge conditions and four ghost fields.

For three dimensional gravity there are several gauges of this type that can be chosen: For example one has:

(i) Diagonal gauge:

Here the 3-metric is of the form:

$$g_{\mu\nu} = \begin{pmatrix} g_{11} & 0 & 0 \\ 0 & g_{22} & 0 \\ 0 & 0 & g_{33} \end{pmatrix} = \begin{pmatrix} e^{\phi_1} & 0 & 0 \\ 0 & e^{\phi_2} & 0 \\ 0 & 0 & e^{\phi_3} \end{pmatrix}$$

(2.3)

(ii) 2d conformal gauge:

With:

$$g_{\mu\nu} = \begin{pmatrix} g_{11} & 0 & g_{13} \\ 0 & g_{11} & g_{23} \\ g_{31} & g_{23} & 1+g_{11}^{-1}(g_{13}^2 + g_{23}^2) \end{pmatrix} = \begin{pmatrix} e^{\phi_1} & 0 & \phi_2 \\ 0 & e^{\phi_1} & \phi_3 \\ \phi_2 & \phi_3 & 1+e^{-\phi_1}(\phi_2^2 + \phi_3^2) \end{pmatrix}$$

(2.4)

(iii) Canonical gauge:

$$g_{\mu\nu} = \begin{pmatrix} h & 0 \\ 0 & 1 \end{pmatrix} = \begin{pmatrix} e^{\phi_1-\phi_2} & e^{\phi_1-\phi_2}\phi_3 & 0 \\ e^{\phi_1-\phi_2}\phi_3 & e^{\phi_1+\phi_2}+e^{\phi_1-\phi_2}\phi_3^2 & 0 \\ 0 & 0 & 1 \end{pmatrix}$$

(2.5)

This is called the canonical gauge because this form of the gauge is often used to write the action in terms of the spatial metric and its conjugate momentum in the canonical formalism. This gauge is also a 3d version of the gauge (2.2) used by E and H. Verlinde [15] for 4d gravity.

The quantization procedure is similar for all three gauges. For definiteness of discussion we choose the canonical gauge. In the canonical gauge the 3d gravitational action becomes:

$$S_{grav} = \int d^3\sigma \sqrt{g} R = \int d^3\sigma \frac{1}{4\sqrt{h_{11}h_{22} - h_{12}^2}} (\partial_3 h_{11}\partial_3 h_{22} - \partial_3 h_{12}\partial_3 h_{12})$$

$$= \int d^3\sigma \frac{1}{4} e^{\phi_1}(-\partial_3\phi_1\partial_3\phi_1 + \partial_3\phi_2\partial_3\phi_2 + e^{-2\phi_2}\partial_3\phi_3\partial_3\phi_3)$$

(2.6)

Where we have set $16\pi G_3 = 1$. The ghost action can be obtained from the variation of the gauge conditions $h^{33} = 1, h^{3a} = 0$ under reparametrizations $\xi^\mu$ and one obtains:

$$\delta h^{33} = 2h^{33}\partial_3 \xi^3$$
$$\delta h^{a3} = h^{ab}\partial_b \xi^3 + h^{33}\partial_3 \xi^a$$

(2.7)

So that Faddev-Popov determinants can be expressed as:

$$\det(\delta g^{13}/\delta \xi)\det(\delta g^{23}/\delta \xi)\det(\delta g^{33}/\delta \xi) = \int DbDce^{-S_{ghost}}$$

(2.8)

Where $c, b$ are ghost and antighost fields and one has:

$$S_{ghost} = \int d^3\sigma \sqrt{h_{33}} \sqrt{h}(h^{33}(2b_{33}\partial_3 c^3 + b_{3a}\partial_3 c^a) + h^{ab}(b_{a3}\partial_b c^3))$$

(2.9)

As the canonical gauge (2.6) is a three dimension version of the gauge used by E. and H. Verlinde [15] to study high energy scattering in 4d gravity it is worth making a comparison. In their case they use:

$$g_{\mu\nu} = \begin{pmatrix} h_{ab} & 0 \\ 0 & h'_{ij} \end{pmatrix}$$ with gravity and ghost action given by:

$$S_{4dgrav} = \frac{1}{4}\int \sqrt{h'}\sqrt{h}(h^{ij}\partial_i h_{ab}\partial_j h_{cd}\frac{1}{\sqrt{h}}\varepsilon^{ac}\frac{1}{\sqrt{h}}\varepsilon^{bd} + h^{ab}\partial_a h'_{ij}\partial_b h'_{kl}\frac{1}{\sqrt{h'}}\varepsilon^{ik}\frac{1}{\sqrt{h'}}\varepsilon^{jl} + 4R_h^{(2)} + 4R_{h'}^{(2)})$$

$$S_{4dghost} = \int \sqrt{h'}\sqrt{h}(h^{ij}b_{ia}\partial_j c^a + h^{ab}b_{ia}\partial_b c^i)$$

(2.10)

In the 3d gravity case we can take the additional condition $h' = h_{33} = 1$ so that the second and third terms in the gravitational action vanish and the last term is a topological invariant on spaces of the form $\Sigma^{(2)} \times R$.

The basic point of is that for 3d gravity one has three Liouville fields and three ghost fields left over after fixing 3d reparametrizations for the gravitational action. This has important implications for the dimension of the Target space in a 3d gravity approach to M-theory. For example if one starts with eight matter fields coupled to gravity one will obtain three more target space dimensions from the Liouville fields alone. This is different from the 2d gravity approach to string theory where one can obtain only a single extra dimension from the Liouville field. We call the metric fields left over after gauge fixing of the type discussed above Liouville fields to make contact with the terminology used in the 2d gravity approach to string theory. Unlike the 2d gravity case the 3d Liouville field interactions are nontrivial after imposing the gauge conditions on the Einstein-Hilbert action. In 2d gravity one has to go to the quantum level to generate nontrivial interaction.

Two modifications to this gauge fixing procedure occur for fermionic matter fields and for three dimensional manifolds of nontrivial topology. For fermionic matter fields one uses the dreibein $e_\mu^a$ which is related to the metric by $g_{\mu\nu} = e_\mu^a e_\nu^a$. However after one fixes local $SO(3)$ invariance, one has the same counting of fields as the metric tensor. For 3d gravity in the dreibein formalism three Liouville modes are also obtained. One starts with nine components of the dreibein $e_\alpha^i$ fixes the three Local Lorentz invariances (say by choosing $e$ symmetric or lower diagonal) and then the three general coordinate invariances (say by choosing $e_3^i = 0$) for a total $9 - 3_{LL} - 3_{GC} = 3$ remaining Liouville modes. For comparison the 2d gravity in zweibein formalism has been discussed by E. and H. Verlinde [16]. In this case one starts with four zweibein $e_\alpha^a$ fixes Local lorentz invariance, (say by choosing $e$ symmetric) and then fixes two general coordinate invariances (say by setting $e_2^1 = 0, e_1^1 = e_2^2$) to obtain $4 - 1_{LL} - 2_{GC} = 1$ Liouville mode. Also for three dimensional manifolds of nontrivial topology it is not possible to gauge away nondiagonal portions of the metric and one must take into account the modular parameters of the manifold. This is similar to the fact that one cannot gauge away constant gauge fields or Wilson lines on spaces with noncontractible loops We illustrate this for the case of the three torus in the next section when we examine $SL(3,Z)$ invariance.

## III SL(3,Z) invariance of 3d gravity and M-theory

The partition function on a torus gives important information about the structure of the target space theory. For example modular invariance or $SL(2,Z)$ invariance of the two torus amplitude is a consistency condition in string theory leading to anomaly cancellation and determination of the gauge group [17]. Also the spectrum of the theory as well as the correct projections or sum over spin structures can also be obtained from the amplitude [18,19]. The purpose of this paper is to investigate a 3d gravity approach to M-theory so we study the partition function on the three torus $T^3$ given by:

$$Z(T^3) = \int De D\omega D\chi DX D\psi e^{iS_{T^3}(e,\omega,\chi,X,\psi)}$$

(3.1)

Here e, $\omega$, $\chi$ are dreibein, spin connection and Rarita-Schwinger field, and X and $\psi$ are bosonic and fermionic matter fields. The action is given by:

$$S = \int d^3\sigma [\det(e) R - i\varepsilon^{mnp} \bar{\chi}_m D_n \chi_p$$
$$- \det(e)\{\frac{1}{2} g^{mn} \partial_m X^M \partial_n X^M + \bar{\psi}^M \gamma^m \nabla_m \psi^M + \frac{i}{2} \partial_n X^M \bar{\psi}^M \gamma^m \gamma^n \chi_m + \frac{1}{8} \bar{\chi}_m \gamma^n \gamma^m \psi^M \bar{\chi}_n \psi^M \}]$$

(3.2)

This form of $N = 1$ Poincaire supergravity with matter action has been developed in [20] as a locally supersymmetric action in three dimensions or as an action for a spinning membrane. Various generalizations to 3d deSitter supergravity with cosmological constant, conformal supergravity and extended supergravity are possible and summarized in [21]. The theory is invariant under reparametrizations, local rotation and Lorentz transformations and local supersymmetry transformations given by:

$$\delta e_\mu^a = -\partial_\mu \xi^a + \varepsilon^{abc}(\theta_b e_{\mu c} + \xi_b \omega_{\mu c}) + \frac{i}{2}\bar{\varepsilon}\gamma^a \chi_\mu$$

$$\delta \omega_\mu^a = -\partial_\mu \theta^a + \varepsilon^{abc}\theta_b \omega_{\mu c}$$

$$\delta \chi_\mu^\alpha = -\partial_\mu \varepsilon^\alpha - \frac{i}{2}\{\theta^a(\gamma_a \chi_\mu)^\alpha - \omega_\mu^a(\gamma_a \varepsilon)^\alpha\}$$

$$\delta X^M = i\bar{\varepsilon}\psi^M \quad \delta\psi^M = \partial X^M \varepsilon - \frac{i}{2}\bar{\chi}\psi^M \varepsilon$$

(3.3)

In the Chern-Simons gauge theory formulation we have:

$$A_\mu = e_\mu^a P_a + \omega_\mu^a J_a + \bar{\chi}_\mu Q$$

$$u = \xi^a P_a + \theta^a J_a + \bar{\varepsilon}Q$$

(3.4)

Where $A$ is a gauge connection and $u$ is a gauge transformation parameter, $\xi, \theta, \varepsilon$ are parameters associated with reparametrizations, Local Lorentz transformations and local supersymmetry transformations respectively. $P, J, Q$ are the generators of SuperPoincaire group ISO(2,1). The Chern-Simons formulation is the most powerful description for nonperturbative calculations in 3d gravity and we shall return to it in a later section. In this section we will be concerned with the leading order perturbative contribution as it is easier to compare to the string derivation of modular invariance.

An arbitrary variation in $e_\mu^a$ is given by:

$$\delta e_\mu^a = e_\mu^a \delta\lambda - \partial_\mu \xi^a + \varepsilon^{abc}(\rho_b \omega_{\mu c}) + e_{\mu,\tau_i}^a \delta\tau_i$$

(3.5)

where $\lambda, \xi, \tau$ are scale, reparametrization and variations in the modular parameters describing the three torus. The functional measure can be written as:

$$De = D\lambda D\xi^a d^5\tau J(\lambda, \tau)$$

(3.6)

where $J(\lambda, \tau)$ is a Jacobian for the transformation (3.5). So the partition function becomes:

$$Z(T^3) = \int_F d^5\tau \int D\lambda D\xi^a J(\lambda, \tau) e^{-S(\tau, \phi, \rho)} = \int_F d^5\tau Z(\tau)$$

(3.7)

The form (3.7) results after fixing local Lorentz (LL) invariance and factoring out the volume of the scale and reparametrization groups. LL invariance can be fixed for example by restricting integration of the dreibein to lower triagonal matrices. The subscript F indicates that the integration is restricted to the fundamental domain of $SL(3,Z)$.

**(a)  one-loop approximation: gravity contribution**

To examine $SL(3,Z)$ invariance we consider the one-loop approximation where all fields are expanded out to quadratic order. The quadratic expansion can be done in any dimensional gravity and supergravity. However in two and three dimensions the one-loop approximation can take on more significance, for example the conformal dimension of fields in 2d gravity can be calculated this way [1] and in 3d gravity the one-loop approximation is the main contribution in the $(e,\omega)=0$ expansion [10]. In the quadratic approximation to 3d gravity and matter the contributions from the various fields to the partition function factorize and we obtain:

$$Z^{one-loop}(\tau) = Z_{grav}(\tau) Z_{Rarita-Schwinger}(\tau) Z_{matter}(\tau)$$

(3.8)

The gravitation contribution $Z_{grav}(\tau)$ is independent of the matter fields as we are working to one-loop. Therefore it must be the same as in the pure gravity case. As with any quadratic path integral it can be expressed as a determinant and takes the general form:

$$Z_{grav}(\tau) = \int DhDbDce^{-S^\tau_{grav}-S^\tau_{ghost}} = \mu_{zero}(\tau) e^{C(\tau)} Z_{osc}(\tau)$$

(3.9)

where the gauge fixed action is specified at particular modular parameter $\tau$. The first factor $\mu_{zero}(\tau)$ is the contribution of zero modes, $C(\tau)$ is the Casimir energy and $Z_{osc}(\tau)$ is the particle or oscillator contribution.

In three dimensions there are no gravitons so that the gravitational contribution to the Casimir energy is zero and the oscillator contribution is unity. This means that in three dimensions the gravitational contribution only comes from the zero modes in the one-loop approximation. This can be further understood by examining the Chern-Simons [10] or phase space path integral [11] or covariant path integral [22]. This is somewhat easier than the two dimensional gravity case which requires an additional scalar field to achieve exactly zero field theoretic degrees of freedom [23]. Note that it is only the product of the Liouville and ghost partition functions that is absent of oscillating terms, each term separately contains them so that the ghosts effectively cancel oscillating terms induced from the three Liouville modes.

This cancellation of determinants has been demonstrated explicitly by Dasgupta and Loll [24] who showed that the path integral in the quadratic approximation is given by:

$$Z^{(2)}(T^3) = \int Dh D\lambda J_1 e^{-S^{(2)}(\tilde{h},\lambda)} = \int D\xi Dm D\lambda J_1 J_2 e^{-S^{(2)}(\xi,m,\lambda)} = J_1 J_2 Z_1 Z_2 Z_3$$

(3.10)

where they define:

$$J_1 = (\det{}^{(1)}(F \circ F^\dagger)^{-1}(F \circ L)^\dagger(F \circ L))^{1/2}$$
$$J_2 = (\det{}^{(1)}(\tilde{L}^\dagger \tilde{L}))^{1/2} \det\langle \chi^{(\alpha)}, \Psi_{(\beta)} \rangle$$

(3.11)

and

$$(L\xi)_{\alpha\beta} = \nabla_\alpha \xi_\beta + \nabla_\beta \xi_\alpha$$
$$(F^\dagger \xi)_{\alpha\beta} = \frac{1}{2}(\xi_\alpha g_{\beta 3} + \xi_\beta g_{\alpha 3}) - \frac{1}{4}\xi_3 g_{\alpha\beta}$$
$$(\tilde{L}\xi)_{ab} = \nabla_a \xi_b + \nabla_b \xi_a - g_{ab}\nabla_c \xi^c \quad \text{for} \quad a,b=1,2$$

(3.12)

In this expression $J_1$ is the Jacobian arising going from the three metric to scale factor $\lambda$ and spatial metric $h$ associated with the canonical gauge, $J_2$ is the Jacobian associated with trading the spatial metric for two reparametrizations $\xi$ and modular fields $m$ associated with the transformation: $\delta h_{ab} = (\tilde{L}\xi)_{ab} + \delta m_\alpha \langle \chi^{(\alpha)} \Psi_{(\beta)} \rangle \delta^{\beta\gamma} \Psi_{(\gamma)ab}$. Here $\Psi_{(\alpha)ab}$ form a basis of $\ker(\tilde{L}^\dagger)$ and $\chi_{ab}^{(\alpha)} = \frac{\partial \hat{g}_{ab}(\tau)}{\partial \tau_\alpha}$. The factors $Z_1, Z_2, Z_3$ determinants arising from the integration with respect to $\lambda, \xi, m$ of the quadratic action and all determinants are evaluated at $\hat{g}(\tau)$. One also has to factor out the infinite volume of Killing vectors on the torus from $\det(\tilde{L}^\dagger \tilde{L})$. The results of [24] indicate that all oscillating factors of the Jacobians $J_1, J_2$ cancel against $Z_1, Z_2, Z_3$ for the three torus and one is left with a zero mode contribution to the measure of moduli space $\mu(\tau)$.

The other portion of the gravity partition function comes from the Rarita – Schwinger action.

$$Z_{RS}^{(2)}(T^3) = \int D\bar{\chi}D\chi e^{iS_{RS}^{(2)}(\chi,\omega,X,\psi)} = \int D\bar{\chi}D\chi e^{i\int d^3\sigma \varepsilon^{\alpha\beta\varsigma}\frac{1}{4}\bar{\chi}_\alpha(\partial_\beta + \frac{1}{2}\omega_\beta^a \gamma_a)\chi_\varsigma}$$

(3.13)

The Rarita-Schwinger path integral can be evaluated in the super Chern-Simons formalism [25] or using a supersymmetric version of the Jacobian method above generalizing the 2d supergravity computation in [26]. Although quadratic in fermi fields like the Dirac action the Rarita-Schwinger action has the gauge symmetry $\delta \chi_\mu = D_\mu \varepsilon$ that leads to the fact the that the Rarita-Schwinger action has zero field theoretic degrees of freedom in three dimensions. This means that there is no oscillator or Casimir contribution from $Z_{RS}^{(2)}$. Also the zero mode contribution are proportional to zero modes of the Rarita-Schwinger field which are supermoduli.

Nevertheless the $\omega$ dependence of (3.13) leads to a dependence on spin structure that has applications to the structure of the theory as well as the spectrum. In string theory the spin structure essentially defines the theory and we expect the same to be true here in a three dimensional context. Note that $F_{\mu\nu}^\alpha = (D_\mu \chi_\nu - D_\mu \chi_\nu)^\alpha$ is the Fermi curvature so we proceed as in [10,27] to perform the path integral. For a spinor valued form $\chi^\alpha$ the covariant exterior derivative is $D_\varpi \chi^\alpha = d\chi^\alpha + \frac{1}{2}\varpi^i (\rho_i \chi)^\alpha$. The Rarita-Schwinger action can be gauge fixed by adding:

$$S_{gf} = \int u_\alpha (D_\varpi * \bar\chi)^\alpha + v_\alpha (D_\varpi * \chi)^\alpha$$ with Faddev-Popov determinant $\left|\det \Delta_\varpi^{(0)}\right|^{-2}$. Then the action can be written as $S_{RS} + S_{gf} = \int \bar\chi (D_\varpi \chi + *D_\varpi u) - v * D_\varpi * \chi$. The integral over $\bar\chi$ and $v$ can be done yielding $\delta(*D_\varpi \chi + D_\varpi u)\delta(D_\varpi * \chi) = \delta(L_-(\chi,u))$ and $L_-(\chi,u) = (*D_\varpi \chi + D_\varpi u) + D_\varpi * \chi$. Noting that we are integrating over fermionic $\chi$ and $u$ the path integral becomes $\dfrac{\left|\det(\Delta_\varpi^{(0)})\right|^{-2}}{\left|\det L_-\right|^{-1}}$. Also the inverse determinants are reminiscent of the 2d gravity calculation [23] where the $(b,c)$ ghosts are cancelled in the $(+,+)$ sector by the Rarita-Schwinger bosonic ghosts $(\beta,\gamma)$ defined by $\left|\det(D_{RS})\right|^{-1} = \int D\beta D\gamma e^{i\int \gamma D_{RS} \beta}$ and the Liouville contribution to the partition function is cancelled by the remaining mode of the 2d Rarita-Schwinger field after gauge fixing.

**(b) parametrization of moduli space**

To proceed further we need an explicit realization for the modular parameters. For a constant three torus the dreibein can be factorized [28] after fixing local Lorentz invariance by:

$$\hat{e}_\mu^a = \begin{pmatrix} L_1 & 0 & 0 \\ 0 & L_2 & 0 \\ 0 & 0 & L_3 \end{pmatrix} \begin{pmatrix} 1 & 0 & 0 \\ a_1 & 1 & 0 \\ a_2 & a_3 & 1 \end{pmatrix}$$

(3.14)

The constant three metric is then given by $\hat{g}_{\mu\nu} = \hat{e}_\mu^a \hat{e}_\nu^a$. For the three torus $T^3$ we have five modular parameters which can be expressed as:

$$\tau_1 = \frac{L_1}{L_2} a_1, \tau_2 = \frac{L_1}{L_2}, \tau_3 = \frac{L_3}{L_2}, \tau_4 = \frac{L_1}{L_2} a_2, \tau_5 = a_3$$

(3.15)

So that the metric, dreibein and inverse dreibein become:

$$\left(\hat{g}(\tau)\right)_{\mu\nu} = (\tau_2\tau_3)^{-2/3} \begin{pmatrix} \tau_1^2 + \tau_2^2 + \tau_4^2 & \tau_1 + \tau_4\tau_5 & \tau_3\tau_4 \\ \tau_1 + \tau_4\tau_5 & 1 + \tau_5^2 & \tau_3\tau_5 \\ \tau_3\tau_4 & \tau_3\tau_5 & \tau_3^2 \end{pmatrix}$$

$$(e(\tau))_\alpha^a = (\tau_2\tau_3)^{-1/3} \begin{pmatrix} \tau_2 & 0 & 0 \\ \tau_1 & 1 & 0 \\ \tau_4 & \tau_5 & \tau_3 \end{pmatrix}$$

$$(e^{-1}(\tau))_a^\alpha = (\tau_2\tau_3)^{1/3} \begin{pmatrix} \tau_2^{-1} & 0 & 0 \\ -\tau_1\tau_2^{-1} & 1 & 0 \\ \tau_1\tau_2^{-1}\tau_3^{-1}\tau_5 - \tau_4\tau_2^{-1}\tau_3^{-1} & -\tau_5\tau_3^{-1} & \tau_3^{-1} \end{pmatrix}$$

(3.16)

where we have divided by an overall scale $(L_1^2 L_2^2 L_3^2)^{1/3}$ as the moduli $\tau$ are used to describe the shape of the torus.

As mentioned in section II one has to modify the canonical gauge on spacetimes with noncontractible loops because off diagonal elements can not be gauged away. One way to modify the gauge is define

$$h = (\tau_2\tau_3)^{-2/3} \begin{pmatrix} e^{\phi_1-\phi_2}\tau_3^2 & e^{\phi_1-\phi_2}\tau_3(\phi_3+\tau_5) \\ e^{\phi_1-\phi_2}\tau_3(\phi_3+\tau_5) & e^{\phi_1+\phi_2} + e^{\phi_1-\phi_2}(\phi_3+\tau_5)^2 \end{pmatrix}$$ with off diagonal components

$g_{13} = N_1 = (\tau_2\tau_3)^{-2/3}(\tau_1 + \tau_4\tau_5)$, $g_{23} = N_2 = (\tau_2\tau_3)^{-2/3}\tau_3\tau_5$ and $g_{33} = (\tau_2\tau_3)^{-2/3}\tau_3^2 + h^{ab}N_aN_b$.
After exchange of rows and columns this form reduces to $\hat{g}(\tau)$ as $\phi_{(i)} \to 0$. Using the expression above in the Einstein-Hilbert action expanded to quadratic order we can determine the contribution from the zero modes for the gravitational sector in the one-loop approximation. The zero mode contribution to the quadratic approximation of the gravitational path integral is then:

$$Z_{grav}(\tau) = \mu_{zero}(\tau) = \frac{1}{\tau_2}((\tau_2\tau_3)^{-1/3})^3$$

(3.17)

where the power of three in the final factor comes from the three Liouville modes.

### (c) one-loop approximation- matter contribution

The remaining contribution to the one-loop partition function comes from the matter fields $Z_{matter}(\tau)$. This can be determined to quadratic order in terms of determinants with zeta function regularization [29] or in an oscillator formalism [30] as all quadratic actions with periodic boundary conditions can be expressed as a discrete product of harmonic oscillators. To quadratic order the matter action becomes:

$$S_{matter} = \int d^3\sigma \det(\hat{e}(\tau))\{\frac{1}{2}\hat{g}^{\alpha\beta}(\tau)\partial_\alpha X^M \partial_\beta X^M + \bar{\psi}^M \gamma^\alpha \nabla_\alpha \psi^M\}$$

(3.18)

The path integral over the matter fields then reduces to:

$$Z_{matter}(\tau) = \int DXD\psi e^{-S_{matter}} = (\det_b(\hat{g}^{\alpha\beta}(\tau)\partial_\alpha \partial_\beta))^{-8/2}(\det_f(\hat{e}^\alpha_a(\tau)\gamma^a\partial_\alpha))^{8/2}$$

(3.19)

These determinants on $T^3$ can be calculated using zeta function regularization in a manner very similar to the calculation of Polchinski on $T^2$ [29] or Dolan and Nappi [31] on $T^6$. Consider the bosonic determinant $\det_b(\hat{g}^{\alpha\beta}(\tau)\partial_\alpha\partial_\beta)$. Taking the logarithm and introducing a set of eigenfunctions $e^{2\pi i n \cdot \sigma}$ the oscillator portion of the determinant is given by the $s \to 0$ limit of:

$$-\frac{d}{ds}\log(\det \Delta^{(0)})^{-s} = -\frac{d}{ds}\sum_{(n)}((n_1 - n_2\tau_1 + n_3(\tau_1\tau_5\tau_3^{-1} - \tau_4\tau_3^{-1}))^2 + (n_2\tau_2 - n_3\tau_2\tau_5\tau_3^{-1})^2 + (n_3\tau_2\tau_3^{-1})^2)^{-s}$$

$$= -\frac{d}{ds}\oint dz \sum_{(n_2,n_3)} \frac{e^{i\pi z}}{2i\sin\pi z}((z - n_2\tau_1 + n_3\tau_1\tau_5\tau_3^{-1})^2 + (n_2\tau_2 - n_3\tau_2\tau_5\tau_3^{-1})^2 + (n_3\tau_2\tau_3^{-1})^2)^{-s} + c.c$$

$$= \log(\prod_{n_2,n_3}(1 - \exp(-2\pi\tau_2\sqrt{n_2^2 + 2n_2n_3\frac{\tau_5}{\tau_3} + n_3^2\frac{1+\tau_5^2}{\tau_3^2}}) + 2\pi i\tau_1 n_2 + 2\pi i(\frac{\tau_1\tau_5 - \tau_4}{\tau_3})n_3))^{-1} + c.c$$

$$\log(\prod_{n_2,n_3}(1 - \exp(-2\pi\tau_2\sqrt{g^{ab}_{(2)}(\tau_3,\tau_5)n_a n_b}) + 2\pi i\tau_1 n_2 + 2\pi i(\frac{\tau_1\tau_5 - \tau_4}{\tau_3})n_3))^{-1} + c.c$$

$$= \sum_{(n_2,n_3)}\log(1 - q^{(n_2,n_3)})^{-1} + c.c = \log f_{boson}(\tau) + c.c$$

(3.20)

Where we have used the simplifying notation:

$$q^{(n_2,n_3)} = \exp(-2\pi\tau_2\sqrt{n_2^2 + 2n_2n_3\frac{\tau_5}{\tau_3} + n_3^2\frac{1+\tau_5^2}{\tau_3^2}} + 2\pi i\tau_1 n_2 + 2\pi i(\frac{\tau_1\tau_5 - \tau_4}{\tau_3})n_3)$$

(3.21)

and have defined: $g^{ab}_{(2)}(\tau_3,\tau_5) = \tau_3^{-1}\begin{pmatrix} \tau_3 & \tau_5 \\ \tau_5 & 1+\tau_5^2 \end{pmatrix}$.

The Casmir and zero mode portions of the boson determinant can be determined similarly.

### (d) Sum over spin structures on $T^3$

The fermion deteminants depend on the spin structure which for periodic fields restrict their Fourier expansion on $T^3$. For example if the fermionic field expansion in written terms of $e^{2\pi i(n_1\sigma^1+n_2\sigma^2+n_3\sigma^3)}$ the spin structure restricts $n_i$ to be integral or half integral. For the case where $n_1$ is half integral and $n_2, n_3$ are integral the determinants are evaluated to be:

$$Z_{matter}(\tau) = (\prod_{n_2,n_3}(1-\exp(-2\pi\tau_2\sqrt{n_2^2+2n_2n_3\frac{\tau_5}{\tau_3}+n_3^2\frac{1+\tau_5^2}{\tau_3^3}})+2\pi i\tau_1 n_2+2\pi i(\frac{\tau_1\tau_5-\tau_4}{\tau_3})n_3))^{-8} \cdot$$

$$(\prod_{n_2,n_3}(1+\exp(-2\pi\tau_2\sqrt{n_2^2+2n_2n_3\frac{\tau_5}{\tau_3}+n_3^2\frac{1+\tau_5^2}{\tau_3^3}})+2\pi i\tau_1 n_2+2\pi i(\frac{\tau_1\tau_5-\tau_4}{\tau_3})n_3))^8$$

$$= \prod_{(n_2,n_3)}(1-q^{(n_2,n_3)})^{-8}(1+q^{(n_2,n_3)})^8$$

(3.22)

However this is only one contribution in the sum over spin structures. For $n_3 \neq 0$ there eight possible terms coming the choice of $n_2, n_3$ integral or half-integral. For $n_3 = 0$ we generate sixteen more possibilities in the product coming from the separate spin structures for left and right movers associated with positive and negative $n_2$ [18]. Finally for $n_2, n_3$ both zero we have two possibilities depending on the whether or not one takes a symmetric or asymmetric combination of with respect to the zero mode fermionic projection operator. As we are in odd dimensional Target space (3 Liouville + 8 matter) as well as odd dimensional world-volume space $T^3$ there are no chiral fermions in the theory in either sense. Thus and we choose the symmetric combination of zero mode projections as is consistent with reduction to *Type IIA* string theory. For manifolds with world volume boundaries and boundaries in Target space there is the possibility of chiral fermions as was shown by Horava and Witten [4] and we shall discuss that case in the next section. We can further reduce the allowed combination of spin structures form the considerations of Dolan and Nappi [31]. In [31] Dolan and Nappi considered the $SL(6,Z)$ invariance of six dimensional antisymmetric tensor field on a six torus. They found using the oscillator or path integral formalism that the possible violation of $SL(6,Z)$ invariance could not come from non-zero $n_3, n_4, n_5, n_6$ as these are effective massive contributions from the point of view of an effective two dimensional theory. We can apply this analysis to the three dimensional theory considered here so that possible violations of $SL(3,Z)$ come from $n_3 = 0$. The main difference is that we from [31] is that we are dealing with a theory that contains fermions and nontrivial spin structure. This means so that we must choose a combination of the sixteen two dimensional spin structures that is modular invariance in the two dimensional sense. In doing so one reduces the possible spin structures drastically and we are left with a sum of partition functions of the form:

$$Z_{matter}(\tau) = (\tau_2\tau_3)^{-8/3} e^{C_{Boson}(\tau)} |f_{Boson}(\tau)|^{16} \cdot$$
$$2^8 e^{C_{(R-,R-)}(\tau)} \prod_{(n_2,n_3)} (1+q^{(n_2,n_3)})^8 (1+q^{(n_2,n_3)})^{8*}$$
$$+ e^{C_{(NS-,NS-)}(\tau)} \cdot \prod_{(n_2,n_3)} (1+q^{(n_2+1/2,n_3)})^8 (1+q^{(n_2+1/2,n_3)})^{8*}$$
$$+ e^{C_{(NS+,NS+)}(\tau)} \prod_{(n_2,n_3)} (1-q^{(n_2+1/2,n_3)})^8 (1-q^{(n_2+1/2,n_3)})^{8*}$$
$$- 2^4 e^{C_{(R-,NS-)}(\tau)} \prod_{(n_2,n_3)} (1+q^{(n_2,n_3)})^8 (1+q^{(n_2+1/2,n_3)})^{8*}$$
$$2^4 e^{C_{(R-,NS+)}(\tau)} \prod_{(n_2>0,n_3=0)} (1+q^{(n_2,n_3)})^8 (1-q^{(n_2+1/2,n_3)})^{8*}$$
$$- 2^4 e^{C_{(NS-,R-)}(\tau)} \prod_{(n_2>0,n_3=0)} (1+q^{(n_2+1/2,n_3)})^8 (1+q^{(n_2,n_3)})^{8*}$$
$$+ 2^4 e^{C_{(NS+,R-)}(\tau)} \prod_{(n_2>0,n_3=0)} (1-q^{(n_2+1/2,n_3)})^8 (1+q^{(n_2,n_3)})^{8*}$$
$$- 2^4 e^{C_{(NS-,NS+)}(\tau)} \prod_{(n_2>0,n_3=0)} (1+q^{(n_2+1/2,n_3)})^8 (1-q^{(n_2+1/2,n_3)})^{8*}$$
$$- 2^4 e^{C_{(NS+,NS-)}(\tau)} \prod_{(n_2>0,n_3=0)} (1-q^{(n_2+1/2,n_3)})^8 (1+q^{(n_2+1/2,n_3)})^{8*}$$

(3.23)

Where we have defined the spin structure dependent Casimir energies [32]:
$C_{(\sigma)}(\tau) = \tau_2 C_{(\sigma)}^{(0)} + \tau_2 C_{(\sigma)}^{(n_3 \neq 0)}(\tau_3, \tau_5)$ where:

$$(C_{Boson}^{(0)}, C^{(0)}_{(R\pm,R\pm)}, C^{(0)}_{(NS\pm,R\pm)}, C^{(0)}_{(R\pm,NS\pm)}, C^{(0)}_{(NS\pm,NS\pm)}) = (-\frac{1}{12}, \frac{1}{12}, \frac{1}{48}, \frac{1}{48}, -\frac{1}{24}) \cdot 8$$

(3.24)

and $C_{(\sigma)}^{(n_3 \neq 0)}(\tau) = 8 \cdot \tau_3^{-3/2} \sum_{n_2,n_3 \in Z'_{(\sigma)}} (n_2^2 + 2n_2 n_3 \frac{\tau_5}{\tau_3} + n_3^2 \frac{1+\tau_5^2}{\tau_3^2})^{-3/2}$ .With integer and half-integer choices for $Z'_{(\sigma)}$ depending on the spin structure.

**(e) *SL(3,Z)* invariance on $T^3$**

Now *SL(3,Z)* invariance is generated by two transformations [31,33]:

$$U_1 = \begin{pmatrix} 0 & 1 & 0 \\ 0 & 0 & 1 \\ 1 & 0 & 0 \end{pmatrix} \quad U_2 = \begin{pmatrix} 1 & 0 & 0 \\ 1 & 1 & 0 \\ 0 & 0 & 1 \end{pmatrix}$$

(3.25)

so that *SL(3,Z)* transformations of the modular parameters are induced by

$$\hat{g}_{\mu\nu}(\tau') = (L\hat{g}(\tau)L^T)_{\mu\nu} \text{ where } L = U_1^{n_1} U_2^{n_2} U_1^{n_3} \ldots$$

(3.26)

Modular *SL(3,Z)* invariance of the partition function can either be verified directly using these transformations or indirectly using the method of Dolan and Nappi [31]. In the indirect method one uses manifest *SL(2,Z)* invariance of the under the transformation $g_{(2)ab}(\tau_3', \tau_5') = (M g_{(2)}(\tau_3, \tau_5) M^T)_{ab}$ together *SL(2,Z)* invariance of the piece of the partition function with $n_3 = 0$ which are generated by T: $\tau_1 \to \tau_1 + 1$ and S: $(\tau_1 + i\tau_2) \to -1/(\tau_1 + i\tau_2)$. In our case we have manifest $SL(2,Z)$ $(\tau_3, \tau_5)$ invariance for the expression under the square root in (3.22), and the spin structures were chosen to reduce upon $n_3 = 0$ to the partition function for type IIA string theory, and this is in turn is $SL(2,Z)$ modular invariant with respect to $(\tau_1, \tau_2)$. As these two transformations are enough to generate all of $SL(3,Z)$ we thus have full $SL(3,Z)$ invariance for the 3d gravity approach to M-theory theory described by (3.23).

**(f) M-theory spectrum from one-loop partition function**

A full analysis of the spectrum requires a nonperturbative treatment of the 3d gravity to determine the physical states, still the one-loop approximation can give some indication of the field content. As in string theory the spectrum of states can be determined from the torus amplitude (3.23). From the $n_3 = 0$ sector we the partition function reduces to the partition function of Type IIA by construction and in order to obtain an SL(3,Z) invariant torus amplitude. We have bosonic states (1+35+28) $\Phi, G_{MN}, B_{MN}$ and (8+56) $A_M, A_{MNP}$ from the $(NS+, NS+)$ and $(R, R)$ terms in the partition function respectively [17]. The structure of these states in type IIA string theory comes from a single superLiouville mode $\chi(n_2) = \chi(\pm 1/2); \chi(0)$ and nine fermion fields $\psi^M(n_2), \psi^9(n_2) = \psi^M(\pm 1/2); \psi^M(0); \psi^9(\pm 1/2); \psi^9(0)$. The bosonic target space is described by zero modes of single Liouville $\phi(n_2) = \phi(0)$ and nine bosonic fields $X^M(n_2), X^9(n_2) = X^M(0); X^9(0)$. The spectrum of states for the type IIA string comes about by choosing eight (1+9−2) fields from $(\chi, \psi^M, \psi^9)$ and using the creation operators to create the lowest states. For example $(G_{MN}, B_{MN})$ from $\psi^{\dagger M}(1/2)\psi^{\dagger N}(-1/2)$. Here the choice of (1+9-2) fields is dictated because we are looking for masssless states in a ten dimensional target space, one Liouville and nine matter.

For In the 3d gravity point of view we have the three components of the Rarita-Schwinger field

$\chi_{(i)}(n_2,n_3) = \chi_1(\pm 1/2,0); \chi_1(\pm 1/2,0); \chi_1(0,0); \chi_2(\pm 1/2,0); \chi_2(\pm 1/2,0); \chi_2(0,0); \chi_3(\pm 1/2,0); \chi_3(0,0).$

and eight fermion fields $\psi^M(n_2,n_3) = \psi^M(\pm 1/2,0); \psi^M(0,0)$. The 3d gravity target space is described by three Liouville zero modes $\phi_{(i)}(n_2,n_3) = \phi_1(0,0); \phi_2(0,0); \phi_3(0,0)$ and eight bosonic fields $X^M(n_2,n_3) = X^M(0,0)$. The spectrum of states is determined by choosing nine $(3+8-2)$ fields from $(\chi_1, \chi_2, \chi_3, \psi^M)$ and using creation operators based on them to create the lowest states. For example $G_{\mu\nu} = (G_{22}, G_{2M}, G_{MN})$ from

$(\chi_2^\dagger(1/2,0)\chi_2^\dagger(-1/2,0); \chi_2^\dagger(1/2,0)\psi^{M\dagger}(-1/2,0); \psi^{M\dagger}(1/2,0)\psi^{N\dagger}(-1/2,0))$. Here the choice of (3+8-2) fields is dictated because we are looking for massless states in an eleven dimensional target space, 3 Liouville and eight matter.

Note that at the lowest level we have the same number of states in different representations. The mapping of lowest states in the 2d gravity type IIA string $(\psi^M)$ representation to states in the 3d gravity M-theory $(\chi_2, \psi^M)$ representation is given by:

$$(66, 44, G_{\mu\nu}) = (55, 35, G_{MN}) \oplus (10, 8, A_M) \oplus (1, 1, \Phi)$$
$$(165, 84, A_{\mu\nu\lambda}) = (45, 28, B_{MN}) \oplus (120, 56, A_{MNP})$$
$$(128, \Psi_\mu) = (56+8, \Psi_M^{(1)}) \oplus (56+8, \Psi_M^{(2)})$$

(3.27)

We have used the notation $(c, d, F)$ denotes (number of component fields, number of degrees of freedom, Field content). This is in agreement with the dimensional reduction counting relating eleven dimensional supergravity and type IIA supergravity in ten dimensions discussed in [4].

**(g) Fundamental region of $SL(3,Z)$**

Finally we describe the fundamental region $F$ used in the integral over moduli space of the three torus. For a general parametrization of moduli space with metric $\hat{g}_{\mu\nu}(\tau)$ a fundamental region can be defined by [34]:

$$\hat{g}_{33}(\tau) \geq \hat{g}_{22}(\tau) \quad \hat{g}_{22}(\tau) \geq \hat{g}_{11}(\tau)$$
$$|\hat{g}_{23}(\tau)| \leq \frac{1}{2}\hat{g}_{33}(\tau) \quad |\hat{g}_{13}(\tau)| \leq \frac{1}{2}\hat{g}_{33}(\tau)$$
$$|\hat{g}_{12}(\tau)| \leq \frac{1}{2}\hat{g}_{22}(\tau)$$

(3.28)

Of course one can choose other representations for the moduli parameters than (3.16). For example if one chooses a representation given by:

$$\tau_1 = \frac{L_1}{L_3}n_1, \tau_2 = \frac{L_1}{L_3}, \tau_3 = \frac{L_2}{L_3}, \tau_4 = \frac{L_1}{L_3}n_2, \tau_5 = \frac{L_2}{L_3}n_3$$

$$(\hat{g}(\tau))_{\mu\nu} = (\tau_2\tau_3)^{-2/3}\begin{pmatrix} \tau_1^2 + \tau_2^2 + \tau_4^2 & \tau_1\tau_3 + \tau_4\tau_5 & \tau_4 \\ \tau_1\tau_3 + \tau_4\tau_5 & \tau_3^2 + \tau_5^2 & \tau_5 \\ \tau_4 & \tau_5 & 1 \end{pmatrix}$$

(3.29)

and the fundamental region can be simply described by the relations:

$$\tau_3^2 + \tau_5^2 \geq 1 \quad \tau_1^2 + \tau_2^2 + \tau_4^2 \geq \tau_3^2 + \tau_5^2$$

$$|\tau_4| \leq \frac{1}{2} \quad |\tau_5| \leq \frac{1}{2}$$

$$|\tau_1\tau_3 + \tau_4\tau_5| \leq \frac{1}{2}(\tau_3^2 + \tau_5^2)$$

(3.30)

Note that, as in the case for $SL(2,Z)$, the dangerous ultraviolet region of parameter space $\tau_2 = 0$ is excluded from the Fundamental region of $T^3$ to ensure that there is no overcounting under $SL(3,Z)$.

The 3d gravity description of M-theory does not contain gauge fields or chiral fermions, two elements which are essential in order to extract realistic physics. The results of [4] show that this can be overcome for an odd dimensional theory by considering Manifolds with boundary in Target space and on the world volume. It is this subject that we turn to in the next section from a 3d gravity approach.

## IV 3d gravity with boundary and gauge fields in *M*-theory

To incorporate gauge fields and chiral fermions in the 3d gravity approach to M-theory we follow the 2d gravity approach to open string theory and introduce a three dimensional manifold with boundary [4]. In particular as a three dimensional manifold with boundary we take the three dimensional annulus or cylinder $I \times T^2$ where I is the interval [0,1]. The boundary is then specified by two $T^2$ components 0 and 1. The modular parameters of the three annulus or cylinder are simpler than the three torus because parameters $a_2, a_3$ are zero and we have :

$$\tau_1 = \frac{L_1}{L_2}a_1, \tau_2 = \frac{L_1}{L_2}, \tau_3 = \frac{L_3}{L_2}$$

(4.1)

Here $\tau_1, \tau_2$ describe the shape of the torus and $\tau_3$ describes the height of the cylinder. The three metric of $I \times T^2$ is then written as:

$$(\hat{g}(\tau))_{\mu\nu} = (\tau_2\tau_3)^{-2/3} \begin{pmatrix} \tau_1^2 + \tau_2^2 & \tau_1 & 0 \\ \tau_1 & 1 & 0 \\ 0 & 0 & \tau_3^2 \end{pmatrix}$$

(4.2)

The action is then modified by a boundary term and we have [4,35]

$$S = S_{grav} + S_{ghost} + S_{bulk-matter} + \int_0 d^2\sigma(\frac{1}{4}i\psi_+^M \partial_-\psi_+^M + \frac{1}{2}i\lambda_-^A \partial_+\lambda_-^A) + \int_1 d^2\sigma(\frac{1}{4}i\psi_+^M \partial_-\psi_+^M + \frac{1}{2}i\lambda_-^{'A} \partial_+\lambda_-^{'A})$$

(4.3)

where the last term is the contribution of the boundary action and represents a right moving $O(8)$ and left moving $E_8$ and $E_8'$ current algebra on each boundary. The partition function is given by the 3d gravity path integral:

$$Z(I \times T^2) = \int DeD\omega D\chi DX D\psi D\lambda D\lambda' e^{-S}$$

(4.4)

Using (3.5) for the arbitrary variation of the dreibein and fixing local Lorentz invariance we can write the path integral as:

$$Z(I \times T^2) = \int_F d^3\tau \int D\varphi D\xi^a J(\varphi,\tau)e^{-S(\tau,\varphi,\xi)} = \int_F d^3\tau Z(\tau)$$

(4.5)

Again performing the field expansion to one-loop or quadratic order the path integral factorizes into

$$Z^{(2)}(\tau) = Z_{grav}(\tau)Z_{bulk-matter}(\tau)Z_{bndry-matter}(\tau)$$

(4.6)

with $Z_{grav}(\tau) = \frac{1}{\tau_2}((\tau_2\tau_3)^{-1/3})^3$ and

$$Z_{bulk-matter}(\tau) = -16q^{*1/3}(e^{C_{R-}(\tau)})^* \prod_{n_2>0,n_3=0}(1+q^{(n_2,0)})^{*8} \cdot \prod_{n_2>0,n_3>0}(1+q^{(n_2,n_3)})^{*8}(1+q^{(n_2,-n_3)})^{*8}$$
$$+q^{*-1/6}(e^{C_{NS-}(\tau)})^* \prod_{n_2\geq 0,n_3=0}(1+q^{(n_2+1/2,0)})^{*8} \cdot \prod_{n_2\geq 0,n_3>0}(1+q^{(n_2+1/2,n_3)})^{*8}(1+q^{(n_2+1/2,-n_3)})^{*8}$$
$$-q^{*-1/6}(e^{C_{NS+}(\tau)})^* \prod_{n_2\geq 0,n_3=0}(1-q^{(n_2+1/2,0)})^{*8} \cdot \prod_{n_2\geq 0,n_3>0}(1-q^{(n_2+1/2,n_3)})^{*8}(1-q^{(n_2+1/2,-n_3)})^{*8}$$

(4.7)

We have defined $q^{(n_2,n_3)} = e^{2\pi i n_2 \tau_1 - 2\pi \tau_2 \sqrt{n_2^2 + n_3^2 \tau_3^{-2}}}$ to simplify the expression..

In (4.3) both boundaries are evaluated at the same modular parameter because the annulus represents a scale expansion of a two torus to another with no shape change. To obtain the correct partition function for the Heterotic string for $n_3 = 0$ we follow [4,35] and take separate $E_8$ and $E_8'$ current algebras on each boundary. Then we have $Z_{bndry-matter}(\tau) = (Z_{E_8}(\tau))^2$ where the partition function on $E_8$ is given by [17]:

$$Z_{E_8}(\tau) = 2^{-1} \prod_{n_2>0, n_3=0}(1+q^{(n_2,0)})^{16} + \prod_{n_2\geq 0, n_3=0}(1+q^{(n_2+1/2,0)})^{16} + 2^7 q \prod_{n_2\geq 0, n_3=0}(1-q^{(n_2+1/2,0)})^{16}$$

(4.8)

The modular group of the annulus is simpler than that for the three torus studied in the previous section. This and because of the form of the metric (4.2) indicates that modular group is given by $SL(2,Z)$. Because we have chosen spin structures that reduce to the Heterotic string for $n_3 = 0$ and the Heterotic string is invariant under $SL(2,Z)$ together with the fact that product terms with $n_3 \neq 0$ act like massive modes and so do not cause anomalies in $SL(2,Z)$ the total partition function for the annulus $Z(I \times T^2)$ is modular invariant. However the fundamental region in this case is the usual $(\tau_1^2 + \tau_2^2) \geq 1$; $|\tau_1| \leq \frac{1}{2}$; $\tau_3 \geq 0$. In this case the dangerous region $\tau_3 = 0$ is not excluded. Again this is familiar from open string theory on an annulus [36]. One still has the possibility of including contributions from other 3d manifolds such as $S^1 \times M^2$ and $I \times K^2$ [37] to invoke cancellation of this region, where $M^2$ and $K^2$ are the Mobius strip and Klein bottle. Finally the use of a well defined current algebra for the boundary action (4.3) leads to the possibility of defining interactions for the gauge fields through vertex operators on the boundary which is in this case two dimensional. Again this is analogous to open string theory where one has gauge interactions are generated by vertex operators on a one dimensional boundary.

## V Pure *3d* gravity and *M*-theory analog of c=1 2d gravity model.

In string theory it has been useful to consider lower dimensional models, such as the c=1 2d gravity model with two target space dimensions, which have a higher degree of solvability than their higher dimensional counterparts [38]. It should also therefore be useful to pursue lower dimensional versions of M-theory for which powerful nonperturbative techniques may also be present. In this section we examine a pure 3d gravity approach to M-theory with three target space dimensions. For the pure 3d gravity approach to M-theory we will consider three cases: pure 3d gravity without matter, pure 3d supergravity without matter and finally pure 3d gravity with boundary matter and target space gauge fields.

**(a)　Pure bosonic 3d gravity and M-theory in three target space dimensions**

For 3d gravity without matter we have three Liouville fields from $h$ (2.5) and three ghosts (2.9) from the three dimensional diffeomorphism invariance for a total of zero degrees of freedom. For pure 3d gravity one also has powerful nonperturbative approaches like the Chern-Simons gauge theory [10], phase space path integral [11] or Lattice gravity [12]. Using the expansion (3.5) and parametrization (2.5) the path integral on the torus for 3d gravity to quadratic order has the general form:

$$Z^{(2)}(T^3) = \int DeD\omega e^{-S^{(2)}} = \int DhDbDc e^{-S_{grav}-S_{ghost}} = \int_F d^5\tau Z_{3dgrav}(\tau)$$

(5.1)

As there is no matter and the graviton contribution is absent in 3d gravity the Casimir and oscillator factors are unity and the partition function is determined from the integral over zero modes. For the parametrization (3.16) we have $Z_{3dgrav}(\tau) = \frac{1}{\tau_2}((\tau_2\tau_3)^{-1/3})^3$ where the power of three is indicative of three target space dimensions. In addition the phase space path integral [11] and Chern Simons gauge theory [10] indicate that to all orders the path integral reduces to an integral over moduli space and can be expressed as a determinant which indicates extension of the one loop result.

How does this compare to the 2d gravity c=1 matter model? For 2d gravity coupled to c=1 matter we have one Liouville field and one matter field, together with two ghosts from two dimensional diffeomorphism invariance for a total of zero degrees of freedom. One also has powerful nonperturbative approaches for 2d gravity such as Poincaire gauge theory [16], exact solution to Liouville theory [39] as well as matrix models [40]. For 2d gravity coupled to c=1 matter the path integral on the torus to quadratic order has the form:

$$Z^{(2)}(T^2) = \int DeD\omega DXe^{-S^{(2)}} = \int D\phi DbDcDXe^{-S_{Liouville}-S_{ghost}-S_X} = \int_F d^2\tau Z_{2dgrav}(\tau)$$

(5.2)

Again because there are zero degrees of freedom the Casimir and oscillator factors are unity so that the entire partition function is determined by zero modes. The torus partition function has been worked out in [23] with the result $Z_{2dgrav}(\tau) = \frac{1}{\tau_2}(\tau_2^{-1/2})^2$ with the power of two is indicative of two target space dimensions, one dimension from the Liouville and one dimension from the c=1 matter field. In addition the exact solutions to Liouville theory [39] and the Matrix Model [40] show that the one loop result holds to all orders on genus one.

**(b) Pure 3d supergravity and M-theory in three Target space dimensions**

For 3d supergravity without matter one again has exactly zero field theoretic degrees of freedom. However for 3d supergravity one has nontrivial spin structure of the Rarita-Schwinger field on a nonsimply connected space. For the case of a three torus we again seek a sum over spin structures consistent with $SL(3,Z)$ invariance. To obtain a consistent spin structure for pure 3d supergravity it is somewhat simpler to if one introduces $p$ matter supermultiplets, studies the partition function, and then takes the limit $p \to 0$.

$$Z_{3dsgrav}(T^3) = \int DeD\omega D\chi DX^p D\psi^p e^{-S}$$
$$= \int DhDbDcD\zeta D\beta D\gamma DX^p D\psi^p e^{-S_{sgrav}-S_{ghost}-S_{matter,p}} = \int_F d^5\tau Z_{3dsgrav,p}(\tau)$$
(5.3)

Expanding fields out to quadratic order the path integral factorizes to a product of determinants depending on the spin structure. Performing the sum over spin structures in the form:

$$Z_{3dsgrav,p}(\tau) =$$

$$2^p |q|^{p/12} e^{C_{R-R-}(\tau)} \prod_{n_2>0,n_3=0} \left|(1+q^{(n_2,0)})\right|^{2p} \cdot \prod_{n_2>0,n_3>0} \left|(1+q^{(n_2,n_3)})\right|^{2p} \cdot \prod_{n_2>0,n_3>0} \left|(1+q^{(n_2,-n_3)})\right|^{2p}$$

$$+ |q|^{-p/24} e^{C_{NS-NS-}(\tau)} \prod_{n_2 \geq 0,n_3=0} \left|(1+q^{(n_2+1/2,0)})\right|^{2p} \cdot \prod_{n_2 \geq 0,n_3>0} \left|(1+q^{(n_2+1/2,n_3)})\right|^{2p} \cdot \prod_{n_2>0,n_3>0} \left|(1+q^{(n_2+1/2,-n_3)})\right|^{2p}$$

$$+ |q|^{-p/24} e^{C_{NS+NS+}(\tau)} \prod_{n_2 \geq 0,n_3=0} \left|(1-q^{(n_2+1/2,0)})\right|^{2p} \cdot \prod_{n_2 \geq 0,n_3>0} \left|(1-q^{(n_2+1/2,n_3)})\right|^{2p} \cdot \prod_{n_2 \geq 0,n_3>0} \left|(1-q^{(n_2+1/2,-n_3)})\right|^{2p}$$

(5.4)

For $n_3 = 0$ we have the a $SL(2,Z)$ modular invariant partition function of the Type 0A string with sectors $(NS-,NS-)$, $(NS+,NS+)$, $(R+,R+)$ and $(R-,R-)$. As before invariance under $(\tau_1,\tau_2)$ for $(n_3=0)$ together with implicit $(\tau_3,\tau_5)$ modular invariance $(n_3 \neq 0)$ implies $SL(3,Z)$ invariance for (5.4). We are mainly interested in $p=0$ for which the *Type 0A* target space theory has a massless tachyon. In this case the oscillator contribution to the partition function are absent but one still has the underlying spin structures over $T^3$ that define the target space theory.

Again a full description requires nonperturbative methods such as the Chern-Simons supergravity generalized to nonzero cosmological constant. Still the one-loop approximation can give some information about field content. For the 2d supergravity approach to type 0A string theory [41] one has the partition function on a two torus:

$$Z_{2dsgrav}(T^2) = \int DeD\omega D\chi DX D\psi e^{-S} = \int D\phi DbDcD\chi D\beta D\gamma DX D\psi e^{-S_{SL}-S_{ghost}-S_{matter}} = \int d^2\tau Z_{2dsgrav}(\tau)$$
(5.5)

Where $S_{SL}(\phi,\chi)$ is the superliouville action and $S_{matter}(X,\psi)$ describes the action of single scalar superfield and $S_{2dgrav} = S_{SL} + S_{matter}$ is given by [41]:

$$S_{2dgrav} = \int d^2\sigma (\frac{1}{2\pi}(\partial\phi\bar{\partial}\phi - \partial X \bar{\partial} X + \chi\bar{\partial}\chi + \bar{\chi}\partial\bar{\chi} - \psi\bar{\partial}\psi - \bar{\psi}\partial\bar{\psi}) + i\sqrt{\lambda_2}\bar{\chi}\chi e^{\phi} + \frac{1}{2}\pi\lambda_2 e^{2\phi})$$

(5.6)

with $\lambda_2$ a two dimensional cosmological constant. The target space is described by the zero modes of $(\phi, X)$ and is two dimensional. Operators which couple to the background target space fields are formed from mode combinations of $(\chi,\psi)$. The target space fields of *type 0A* $p=0$ string are generated by the ground state, superLiouville field $\chi$ and matter field $\psi$ and are of the form $(1,1,T)$ from $(NS-,NS-)$ ground state, $(3,-1,G_{MN})$, $(1,0,B_{MN})$, $(1,1,\Phi)$ from $(NS+,NS+)$ $(\chi_{-1/2}\tilde{\chi}_{-1/2}, \chi_{-1/2}\tilde{\psi}_{-1/2}, \psi_{-1/2}\tilde{\chi}_{-1/2}, \psi_{-1/2}\tilde{\psi}_{-1/2})$ sector and $(2,0,A^{(1)}{}_M), (2,0,A^{(2)}{}_M)$ from the $(R,R)$ $(\chi_0\tilde{\chi}_0, \chi_0\tilde{\psi}_0, \psi_0\tilde{\chi}_0, \psi_0\tilde{\psi}_0)$ sector. These fields generate an effective action [41]

$$S_{Type-0A-string} = \int d\phi dX \sqrt{-\det G}(-\partial T \partial T + G^{MN} R_{MN} - \frac{1}{2} H^{MNP} H_{MNP} - \partial\Phi\partial\Phi - \frac{1}{2} F^{(1)2} - \frac{1}{2} F^{(2)2})$$

(5.7)

For the 3d supergravity approach to M-theory one has the pure $N=1$ supergravity action with non-zero cosmological constant given by:

$$S_{3dgrav} = \int d^3\sigma \{\frac{1}{4} e^{\phi_1}(-\partial_3\phi_1\partial_3\phi_1 + \partial_3\phi_2\partial_3\phi_2 + e^{-2\phi_2}\partial_3\phi_3\partial_3\phi_3) + \varepsilon^{ijk}\bar{\chi}_i\partial_j\chi_k + \frac{i}{2}\sqrt{\lambda_3} e^{\phi_1/2}\tilde{e}^a_i \varepsilon^{ijk}\bar{\chi}_j\rho_a\chi_k + 2\lambda_3 e^{\phi_1}\}$$

(5.8)

where: $\tilde{e}^a_i(\phi_2,\phi_3) = e^{-\phi_2/2}\begin{pmatrix} 1 & 0 & 0 \\ \phi_3 & e^{\phi_2} & 0 \\ 0 & 0 & 1 \end{pmatrix}$ and $\lambda_3$ is the three dimensional cosmological constant. In this case the target space is given by the zero modes of $(\phi_1,\phi_2,\phi_3)$ and is three dimensional. Operators that couple to the target space background fields are generated by the ground state and products of 3d Rarita Schwinger fourier components:

$\chi_{(i)}(n_2,n_3) =$
$\chi_1(\pm 1/2, 0); \chi_1(\pm 1/2, 0); \chi_1(0,0); \chi_2(\pm 1/2, 0); \chi_2(\pm 1/2, 0); \chi_2(0,0)\chi_3(\pm 1/2, 0); \chi_3(\pm 1/2, 0); \chi_3(0,0).$

(5.9)

The low lying field structure $n_3 = 0$ and rightmoving $n_2 \geq 0$ and left moving $n_2 < 0$ are of the form $(1,1,T) \oplus (6,0,G_{\mu\nu}) \oplus (3,0,B_{\mu\nu})$ where again $(c,d,F)$ denotes (number of component fields, number of degrees of freedom, Field content). Finally the low lying fields form a three dimensional Target space effective action of the form:

$$S_{M-theory, p=0} = \int d^3\phi(\sqrt{-\det G}(-\partial T\partial T + G^{\mu\nu} R_{\mu\nu} - \frac{1}{2} H^{\mu\nu\lambda} H_{\mu\nu\lambda} + \Lambda)$$

(5.10)

Note that the $H_{\mu\nu\lambda}$ can be interpreted as a connection component with torsion, so that the target space gravity for three dimensional M-theory can itself be put in a Chern-Simons form $S_{3dMtheory-grav} = \int E \wedge (d\Omega + \Omega \wedge \Omega) + \Lambda E \wedge E \wedge E$ where $E, \Omega$ are target space dreibein and connection and $\Lambda$ is the target space cosmological constant. In three dimensions the three form field strength has no degrees of freedom and gives an additional contribution to the cosmological constant similar to the four form in 4d gravity.

The states of the type 0A string are related to the three dimensional M theory effective action by compactification on a circle. In particular

$$(6, 0, G_{\mu\nu}) = (3, -1, G_{MN}) \oplus (2, 0, A_M^{(1)}) \oplus (1, 1, \Phi)$$
$$(3, 0, B_{\mu\nu}) = (1, 0, B_{MN}) \oplus (2, 0, A_M^{(2)})$$
$$(1, 1, T) = (1, 1, T)$$

(5.11)

Thus we see that the target space three dimensional M-theory described by (5.11) is related to the type 0A two dimensional string theory in the same way that eleven dimensional M-theory is related to the type IIA ten dimensional string.

### (c) Pure 3d supergravity with boundary and Target space gauge fields.

The above three dimension M-theory the target space fields only contained gravity $G_{\mu\nu}$ and torsion fields $B_{\mu\nu}$ thus it is of interest to see if gauge fields can be added. We shall introduce gauge fields through a two dimensional boundary and current algebra in the world volume 3d supergravity as was done in section IV. For the pure 3d supergravity on boundary we again take the path integral on $I \times T^2$ and form the path integral:

$$Z_{3dsgrav}(I \times T^2) = \int DeD\omega D\chi D\lambda D\lambda' DX^p D\psi^p e^{-S}$$
$$= \int DhDbDcD\zeta D\beta D\gamma D\lambda D\lambda' DX^p D\psi^p e^{-S_{sgrav}-S_{ghost}-S_{bndry}-S_{matter,p}} = \int_F d^3\tau Z_{3dsgrav,p}(\tau)$$

(5.12)

Here the boundary action is given by (4.3) with a current algebra $E_8$ on boundary component 0 and $SO(8+p)$ on component 1 where $I$ is taken as the interval [0,1]. This particular combination is interesting because it yields a modular invariant partition function for $n_3 = 0$ [42]. In addition we have added p bulk matter multiplets to illustrate spin structures but shall take the limit $p \to 0$. To quadratic order the partition function factorizes and we have:

$$Z_{3dsgrav,p}(\tau) = Z_{E_8}(\tau)Z_b(\tau) \cdot \{$$

$$-2^{p+4}|q|^{p/12}(q^*)^{1/3}e^{C_{R-R-}(\tau)}\prod_{n_2>0,n_3=0}\left|(1+q^{(n_2,0)})\right|^{2p}(1+q^{(n_2,0)})^{8*} \cdot \prod_{n_2>0,n_3>0}\left|(1+q^{(n_2,n_3)})\right|^{2p} \cdot \prod_{n_2>0,n_3>0}\left|(1+q^{(n_2,-n_3)})\right|^{2p}$$

$$+|q|^{-p/24}(q^*)^{-1/6}e^{C_{NS-NS-}(\tau)}\prod_{n_2\geq 0,n_3=0}\left|(1+q^{(n_2+1/2,0)})\right|^{2p}(1+q^{(n_2+1/2,0)})^{8*} \cdot \prod_{n_2\geq 0,n_3>0}\left|(1+q^{(n_2+1/2,n_3)})\right|^{2p} \cdot \prod_{n_2>0,n_3>0}\left|(1+q^{(n_2+1/2,-n_3)})\right|^{2p}$$

$$-|q|^{-(p+4)/24}(q^*)^{-1/6}e^{C_{NS+NS+}(\tau)}\prod_{n_2\geq 0,n_3=0}\left|(1-q^{(n_2+1/2,0)})\right|^{2p}(1-q^{(n_2+1/2,0)})^{8*} \cdot \prod_{n_2\geq 0,n_3>0}\left|(1-q^{(n_2+1/2,n_3)})\right|^{2p} \cdot \prod_{n_2\geq 0,n_3>0}\left|(1-q^{(n_2+1/2,-n_3)})\right|^{2p}\}$$

(5.13)

Now taking the limit $p \to 0$ the partition function becomes:

$$Z_{3dsgrav}(\tau) = \frac{1}{\tau_2}((\tau_2\tau_3)^{-1/3})^3 Z_{E_8}(\tau) \cdot$$

$$\cdot(-2^4(q^*)^{1/3}\prod_n(1+q^{n+1/2})^{*8} - (q^*)^{-1/6}\prod_n(1-q^{n+1/2})^{*8} - (q^*)^{-1/6}\prod_n(1+q^n)^{*8})$$

(5.14)

Low lying states of the theory are associated products of oscillator nodes of the Rarita schwinger field and current algebra. In particular $\chi_{(i)}(\pm 1/2,0)\lambda^{\prime\prime}(\pm 1/2)\lambda^{\prime\prime}(\pm 1/2)$ and $\chi_{(i)}(0,0)\lambda^{'}(0)$ yields gauge particles and fermions transforming under $E_8 \times SO(8)$. These states are identified with the spectrum of the Heterotic $E_8 \times SO(8)$ theory in two target space dimensions discussed in [42]. As we have already have three target space dimensions in M-theory from the three Liouville modes we can couple the Heterotic theory to the three dimensional target space along a two dimensional target space boundary. This is a lower dimensional analog to coupling eleven dimensional M-theory to a Heterotic theory on a ten dimensional boundary [4].

## VI. Nonperturbative 3d gravity and M-theory

Three dimensional gravity has powerful nonperturbative approaches like the Chern-Simons gauge formulation, phase space path integral, lattice gravity, even Matrix theory to name a few. Nonperturbative 2d gravity using Pioncaire gauge gravity, exact solutions to Liouville theory and Matrix theory has important applications to string theory, especially for low dimensional target spaces. Thus it is important to apply nonperturbative methods in a 3d gravity approach to M-theory. In particular some areas that are directly relevant to 3d gravity approach to M-theory are: the reduction of a 3d gravity path integral to an integral over moduli, sum over 3d topologies, expansion about $e = 0$ in the 3d gravity path integral, evaluation of loop correlators, evaluation of topology change and nonperturbative definition of physical states. In compiling this list we are appealing to known areas of fruitful application in nonperturbative 2d gravity to string theory. Our studies in the previous sections indicate that there are two cases depending on whether the world volume theory couples to matter or can be described by pure 3d gravity.

### (a) M-theory in three target space dimensions.

In the previous sections we showed that M-theory in three target space dimensions can be described in the 3d gravity approach by a pure 3d gravity or 3d supergravity without matter. In this case pure 3d gravity or supergravity and nonperturbative methods can be readily applied. This means that one can define the amplitudes of M-theory in three target space dimensions without recourse to expansions around quadratic order. One powerful nonperturbative methods of evaluating $Z(M)$ developed through Chern Simons theory and applied to 3d gravity by Witten in [10]. In this section we attempt to apply those methods to the evaluation of vacuum amplitudes in M-theory in three target space dimensions. In particular in the Chern-Simons gauge approach the torus amplitude can be written:

$$Z(T^3) = \int DeD\omega e^{iS_{grav}} = \int DA e^{iS_{CS}(A)}$$
(6.1)

with $S_{grav} = \int e^a \wedge (d\omega^a + \varepsilon^{abc}\omega^b \wedge \omega^c)$ and $S_{CS} = \int tr(A \wedge (dA + A \wedge A))$

with $A$ is related to $(e,\omega)$ through (3.4). Note that the path integral (6.1) is usually defined in Noneuclidean space. For spacetimes of the form (3.16) it relatively straightforward to go from Euclidean to Noneuclidean spacetimes by omitting the $(\tau_2\tau_3)^{-1/3}$ prefactor in $e$ and replacing $\tau_2$ by $i\tau_2$.

The path integral (6.1) after gauge fixing and integration over ghosts reduces to a integral over the moduli space of flat connections on M. The form of the amplitude has been explicitly worked out by Witten in [10] who found that the path integral reduces to:

$$Z(M) = \int de_{(\lambda)} \int_N \frac{(\det \Delta)^2}{|\det' L_-|} = \int_{Moduli} \frac{(\det \Delta)^2}{|\det' L_-|} = \int_{T^*N} d\tilde{\omega} d\tilde{e} T(\tilde{\omega})$$
(6.2)

Here N is the space of tangent vectors, $e_{(\lambda)}$ are dreibein zero modes, $\Delta$ is the scalar Laplacian, $L_-$ is the differential operator $*D + D*$, $d\tilde{\omega}$ is the integration over moduli space of flat connections, $d\tilde{e}$ is the integral over zero modes of the dreibein. The combination of determinants is called the Ray-Singer torsion and can be expressed as:

$$T(\tilde{\omega}) = \frac{\left|\det \Delta_{\tilde{\omega}}^{(0)}\right|^2}{\left|\det \Delta_{\tilde{\omega}}^{(1)}\right|^{1/2} \left|\det \Delta_{\tilde{\omega}}^{(3)}\right|^{1/2}} = \frac{\left|\det \Delta_{\tilde{\omega}}^{(0)}\right|^{3/2}}{\left|\det \Delta_{\tilde{\omega}}^{(1)}\right|^{1/2}}$$
(6.3)

where we have used the Hodge duality relation between $n$ form and $3-n$ Laplacians in three dimensional space [27]. The reduction of the path integral to determinants is indicative of the fact that for pure 3d gravity the expansion about $(e,\omega) = 0$ the one-loop approximation determines the amplitude. However The presence of determinants does not necessarily indicate that oscillating field theoretic degrees of freedom are present in the path integral. For example on $R^3$ the vector Laplacian cancels three scalar Laplacians and $T=1$ indicating no field theoretic or oscillating modes. Although the expression (6.3) reduces the 3d gravity approach to M-theory in three target space dimensions to an integral over moduli space, the expression for $T(\tilde{\omega})$ can be quite complicated. For example an explicit expression for the amplitude on connected components of $\Sigma \times R$ for genus three and two Reimann surfaces $\Sigma$ was carried out in [43]. This is consistent with the 2d gravity approach to string theory where the moduli space of high genus surfaces is also quite complicated.

For three dimensional M-theory with two dimensional boundary matter as in section V the amplitudes can also be expressed in terms of nonperturbative 3d gravity. In this case one has the Heterotic $E_8$ and $D_4$ current algebras on separate boundaries discussed in section V so that the vacuum amplitude is of the form:

$$Z(M) = \int DeD\omega D\psi e^{iS_{3dgrav}(e,\omega) + i\int_\Sigma L_{bndrymatter}(e,\psi) + i\int_{\Sigma'} L_{bndrymatter}(e,\psi')}$$

(6.4)

where the interaction of the 3d gravity with the boundary matter is given by:

$$e\bar{\psi}\gamma^a (e^{-1})_a^\alpha \nabla_\alpha \psi = -\bar{\psi}\gamma^a \varepsilon^{\alpha\beta} e_{\beta b} \varepsilon^{ab} \partial_\alpha \psi$$

(6.5)

Here all indices on the boundary are two dimensional and $e_\alpha^a$ is the dreibein evaluated on the boundary so it becomes an effective zweibein used in the 2d gravity topological approach of [16]. This theory is similar to the topological membrane approach to string theory discussed in [8] where all the matter is contained on a two dimensional boundary. The expression (6.4) for the vacuum amplitude $Z(M)$ written the partition function is related to kernel function $K$ defined in [10] to discuss topology changing transitions:

$$K(\omega|_\Sigma, \omega|_{\Sigma'}) = \int_{\omega=\omega|_\Sigma}^{\omega=\omega|_{\Sigma'}} DeD\omega D\psi e^{iS_{3gdrav}(e,\omega) + iS_{bndry}(e,\psi)}$$

(6.6)

and this is a generalization of the relation in string theory between say the one loop partition function on the annulus and the propagator interpretation of the cylinder.

In our perturbative calculation of the vacuum amplitude on the 3d annulus using the quadratic expansion we found the partition function had factors for the boundary components $Z_{D_4}(\tau)$ and $Z_{E_8}(\tau)$. In that case both boundary components were evaluated at the same 2d boundary moduli $\tau$ because the annulus parameters changed the size but not the shape of the bounding two tori. This has a nonperturbative generalization in Chern-Simons gravity theory. In that case a nonzero transition amplitude $K$ is only obtained if the holonomies on $\Sigma$ and $\Sigma'$ coincide [10,27]. These holonomies are in turn related to the modular parameters through the results of Carlip in [44]. Finally wave functions and hence state spectra can also be discussed in the nonperturbative Chern-Simons formulation of quantum gravity by considering path integrals with a single boundary [45]. The case of three dimensional M-theory is especially interesting because of the correspondence relates the M-theory spectrum on $AdS_3$ to a two dimensional boundary CFT. This fact can also be used on the three dimensional world volume theory to relate the wave functions to two dimensional CFT partition functions as was done in [46].

**(b)  M-theory in higher Target space dimensions and 3d gravity with matter**

For target space dimensions above three it much more challenging to formulate a nonperturbative 3d gravity approach to M-theory. This is because if one formulates the matter contribution coupling to the bulk world volume manifold $M$ and defines vacuum amplitude through:

$$Z(M) = \int DeD\omega DXD\psi e^{i(S_{grav}+S_{matter})}$$

(6.7)

with 3d gravity matter interaction given by (for simplicity we set the Rarita-Schwinger term to zero):

$$S_{matter} = \int d^3\sigma e((e^{-1})_\alpha^a (e^{-1})_\beta^a \partial_\alpha X^M \partial_\beta X^M + (e^{-1})_a^\alpha (\bar\psi^M \gamma^a \nabla_\alpha \psi^M - (\nabla_\alpha \bar\psi^M)\gamma^a \psi^M))$$

(6.8)

Then one has explicit dependence on the inverse dreibein $(e^{-1})_\alpha^a$ and thus it is difficult to form the short distance expansion about $(e,\omega)=0$. Note that in any dimension the gravity action is quadratic in $\omega$ so that it can be integrated over to obtain a path integral over $e$ alone. However when one does this one obtains explicit $(e^{-1})_\alpha^a$ in the second order form of the gravitational action. What is different about 3d gravity in the gauge formulation that it is linear in $e$ so that it can be integrated out first in favor of $\omega$ except for zero modes $e_{(\lambda)}$. This is crucial in the reduction of the 3d gravity path integral to an integral over moduli space. Thus we must seek a formulation of gravity matter interaction that does not depend on the inverse dreibein $(e^{-1})_a^\alpha$. One method used in the gauge approach to supergravity is to introduce auxiliary fields associated with the matter fields through the action:

$$S_{aux} = \int d^3\sigma (P_M^\alpha P_M^\beta e_\alpha^a e_\beta^a + P_M^\alpha \partial_\alpha X^M e^{1/2} + B_{aM}^* C_M^\alpha e_\alpha^a + (B_{aM}, C_M^\alpha)^* \begin{pmatrix} \gamma^a \psi^M \\ \nabla_\alpha \psi^M \end{pmatrix} e^{1/2} + c.c.$$

(6.9)

Using this first order form of the matter action the partition function becomes:

$$Z(M) = \int DeD\omega DP_M^\alpha DB_{aM} DC_M^\alpha DXD\psi e^{iS_{grav} + i\int d^3\sigma(P_M^\alpha P_M^\beta e_\alpha^a e_\beta^a + P_M^\alpha \partial_\alpha X^M e^{1/2} + B_{aM}^* C_M^\alpha e_\alpha^a + (B_{aM}, C_M^\alpha)^* \begin{pmatrix} \gamma^a \psi^M \\ \nabla_\alpha \psi^M \end{pmatrix} e^{1/2} + c.c.}$$

(6.10)

For the simple form of the matter action that we chose in (6.8) the first order form is linear in the matter fields so that these can be integrated out to obtain:

$$Z(M) = \int DeD\omega DPDBDC \delta(\partial_\alpha P^\alpha) \delta(\gamma^a B_a + \partial_\alpha C^\alpha) e^{i\int eR(\omega) + PPee + BCe}$$

(6.11)

The delta function constraints on the auxiliary fields can be implemented by mode expansion on $M$. For example if $M = T^3$ or $T^2 \times I$ we have the expansion $P^\alpha(\sigma) = \sum_{(n)} P(n) e^{2\pi i n \cdot \sigma}$ and the delta function constraint sets

$$P^1(n) = -(n_2 P^2(n) + n_3 P^3(n))\frac{1}{n_1} \text{ for } n \neq 0.$$

The integrated form of the action (6.11) has terms linear and quadratic in the $e$ field. As in the the pure 3d gravity case we can integrate over $e$ in favor of the $\omega$ field taking into account dreibein zero modes $e_{(\lambda)}$, introduce Faddev–Popov ghost determinant associated with gauge conditions $D_a e^a = D_a \omega^a = 0$ as well as Lagrange multiplier $u$ associated with the gauge condition $D_a e^a$ to obtain:

$$Z(M) =$$
$$\int d\hat{e}(\tau) \int D\omega Du DPDBDC (\det(PP))^{-1/2} (\det(\Delta))^2 \delta(D_a \omega^a) \delta(\partial_\alpha P^\alpha) \delta(\gamma^a B_a + \partial_\alpha C^\alpha) e^{i\int (PP)^{-1}(R+BC+*Du)(R+BC+*Du)}$$

(6.12)

As $P$ is related to target space momentum it is of interest to examine the behavior of (6.12) at various values of $P$. For small $P$ the integral becomes a delta function as is evident from (6.10) so that we obtain the leading term:

$$Z(M) = \int d\hat{e}(\tau) \int D\omega DBDC \delta(D_a \omega^a) \delta(\varepsilon^{\alpha\beta\gamma} R_{\beta\gamma}^a(\omega) + B_a C^\alpha + D_\alpha u^a) + \ldots$$

(6.13)

As the $BC$ term is Grassman valued and can be integrated out so we have a term proportional to the $(e, \omega) = 0$ expansion of pure 3d gravity. For arbitrary $P$ the expression (.) is similar to the form of a string propagator expressed in terms of the mode expansion of target space momentum $P(\sigma) = \sum_n P_n e^{2\pi i(n_1 \sigma^1 + n_2 \sigma^2)}$ associated with $X(\sigma) = \sum_{(n)} X_n e^{2\pi i(n_1 \sigma^1 + n_2 \sigma^2)}$. In any case have written an expression for the path integral for the vacuum amplitude for M-theory that does not require the introduction of the inverse dreibein $(e^{-1})_a^\alpha$ and is suitable for expansion about the unbroken generally covariant state $(e, \omega) = 0$.

In a sense the $\omega$ field can be thought of as an auxiliary field for gravity. This is because the gravitational action can be written:

$$S_{3dgrav} = \int \omega \wedge \omega \wedge e - \omega \wedge de$$

(6.14)

This means that in the small $\omega$ limit the equation of motion for $e$ reduces to:

$de^a = 0 + O(\omega)$ with solution $e_\mu^a = \partial_\mu X^a$ found in [47]. This reduces to the topological gravity form of the metric tensor discussed $g_{\mu\nu} = \eta_{ab} e_\mu^a e_\nu^b = \eta_{ab} \partial_\mu X^a \partial_\nu X^b$ by Verlinde [16].

For the matter auxiliary fields we have:

$$S_{matter} = \int P^{\mu M} P^{\nu M} \eta_{ab} e_\mu^a e_\nu^b - P^{\mu M} \partial_\mu X^M e^{1/2}$$

(6.15)

Now in the matter sector in the small $P$ limit the equation of motion for the matter field reduces to: $\partial_\mu X^M = 0 + O(P)$ with solution $X^M = const$. In any event the target space dimension in the small auxiliary field or topological limit is given by the dimension of $(X^a, X^M)$ which is 3 plus the number of matter fields in agreement with the target space dimension determined by counting Liouville modes plus matter fields in section II.

Finally there is another special feature of three dimensional field theory that is useful in studying the target space duality of M-theory in the 3d gravity approach. If one of the matter fields $X$ takes its target space value on a circle of radius $R$ the scalar field is dual in three dimensions to vector field strength through the relation $R\partial_\alpha X = (1/R)\varepsilon_{\alpha\beta\varsigma} F^{\beta\varsigma}$. Thus rather than the phenomena of vortices appearing in the 2d gravity to duality in string theory as in [48] we have monopoles appearing in the 3d gravity approach to M-theory. In terms of the dual $F$ matter field it still possible to use auxiliary fields to avoid $(e^{-1})_a^\alpha$ terms in the action through the auxiliary field action

$S_{aux} = \int \frac{1}{4} L^{\alpha\beta} L^{\varsigma\delta} g_{\alpha\varsigma} g_{\beta\delta} + L^{\alpha\beta} F_{\alpha\beta} e^{1/2}(1/R)$ with $g_{\alpha\beta} = e_\alpha^a e_{a\beta}$ so that under duality

$e_\alpha^a P^\alpha \leftrightarrow \varepsilon_{abc} e_\alpha^a e_\beta^b L^{\alpha\beta}$ while $R \leftrightarrow 1/R$. This is reminiscent of the exchange Target space momentum and winding number found for duality in the string theory.

**VII Liouville 2d gravity from 3d gravity and the critical dimension in M-theory**

For the 3d gravity approach to M-theory it is important to understand the relation of 3d gravity with 2d Liouville gravity in order to relate M-theory and string theory and ultimately understand how M-theory connects the different string theories. Also the conformal dimension of fields is proportional to the induced 2d Liouville term and this can be used to understand the critical dimension of the theory in a conformal gauge like (2.4). Before M-theory was discovered 3d Chern-Simons theories were already used to connect the conformal block representations of conformal field theories. Thus there were already hints a three dimensional theory was necessary to connect string theories in different backgrounds. Also the origin of the Liouville action in 3d gravity and 2d gravity is very different. In 3d gravity we call the remaining components of the action after gauge fixing three Liouville fields as we discussed in section II. There is no difficulty in obtaining an action for these fields, one simply applies the gauge condition to the Einstein-Hilbert action to obtain a non-trivial interacting field theory. In the 2d Liouville gravity the Einsein-Hilbert action is a topological invariant and one has to work much harder and generate a quantum effective action for the single metric remaining after gauge fixing, the Liouville action, to obtain a notrivial interaction field theory of 2d gravity.

In [49] Alavarez-Gaume and Witten computed the 2d gravitational anomaly by studying the Feynman diagram shown in figure 1 which involves the interaction of two stress energy tensors with fermionic matter. They also showed how the conformal anomaly and 2d Liouville action can also studied from the same diagram. In 3d gravity there are no gravitational anomalies. If one couples to 2d boundary matter then anomalies are possible in 3d gravity however in [4] Horava and Witten give an argument that the gravitation anomalies for $E_8$ and $E_8$'on separate 2d boundaries cancel. Still the origin of the conformal anomaly, Virasoro algebra and Liouville action of 2d gravity from a 3d gravity is of great interest [7] both in Chern-Simons formulation and in interaction with matter. Here we study the interaction of 3d gravity with fermionic matter using the diagram in figure 1. As the diagram was used to show the origin of the Liouville action in 2d gravity we will use the same diagram show the origin of 2d Liouville action in 3d gravity with matter. The calculation is simplified because in three dimensions the interaction between gravity and fermionic matter is finite to one-loop [50].

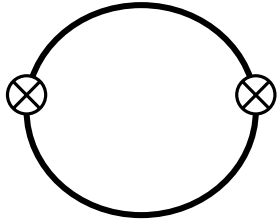

**Figure 1.** Feynman diagram for the 3d gravity interaction of fermionic matter at one loop with two stress energy tensors used to generate an effective gravitational action.

The interaction of fermionic matter with 3d gravity can be described by:

$$\delta L = -\frac{i}{2}\bar{\psi}\gamma^a \varepsilon^{\alpha\beta\rho} e^b_\beta e^c_\rho \varepsilon_{abc} \partial_\alpha \psi = -\frac{i}{8} h^{\alpha\beta}(\bar{\psi}\gamma_\alpha \partial_\beta \psi + \bar{\psi}\gamma_\beta \partial_\alpha \psi - (\partial_\alpha \bar{\psi})\gamma_\beta \psi - (\partial_\beta \bar{\psi})\gamma_\alpha \psi))$$

(7.1)

where we follow the approach of [49] by fixing Local Lorentz invariance by setting $e_\alpha^a = \eta_{a\alpha} + \frac{1}{2}h_{a\alpha}$ with $h_{a\alpha}$ a symmetric metric variation. The second order metric perturbation of figure 1 is then used to used to investigate the quantum effective action. Following the same treatment of [49] however working in three dimensions the amplitude is written:

$$I_{\alpha\beta\varsigma\delta}(p) = -\frac{1}{4}\int\frac{d^3k}{(2\pi)^3}tr(2k+p)_\alpha(\gamma_\beta\frac{1}{(k+p)}(2k+p)_\varsigma\gamma_\delta\frac{1}{(k)})$$

(7.2)

Now using $tr(\gamma_\alpha\gamma_\beta\gamma_\varsigma\gamma_\delta) = (2\eta_{\alpha\beta}\eta_{\varsigma\delta} + 2\eta_{\beta\varsigma}\eta_{\alpha\delta} - 2\eta_{\alpha\varsigma}\eta_{\beta\delta})$, introducing the integration parameter $x$ and shifting loop momentum to $\ell - px$ we have:

$$I_{\alpha\beta\varsigma\delta}(p) = -\frac{1}{4}\int_0^1 dx\int\frac{d^3\ell}{(2\pi)^3}(2\ell - 2px + p)_\alpha(2\ell - 2px + p)_\varsigma \cdot$$
$$\{2(\ell - px + p)_\beta(\ell - px)_\delta + 2(\ell - px + p)_\delta(\ell - px)_\beta - 2\eta_{\beta\delta}(\ell - px + p)\cdot(\ell - px)\}\cdot$$
$$\cdot\frac{1}{(\ell^2 + p^2 x(1-x))^2}$$

(7.3)

Now performing the integral and extracting the piece proportional to four momentum index structure $p_\alpha p_\beta p_\varsigma p_\delta$ denoted by $U_{\alpha\beta\varsigma\delta}(p)$ we have

$$U_{\alpha\beta\varsigma\delta}(p) = \frac{1}{2}p_\alpha p_\beta p_\varsigma p_\delta \int_0^1 dx(1-2x)^2(1-x)x\int\frac{d^3\ell}{(2\pi)^3}\frac{1}{(\ell^2 + x(1-x)p^2)^2}$$

(7.4)

Performing the integral over loop momentum $\ell$ and integration parameter $x$ we obtain:

$$U_{\alpha\beta\varsigma\delta}(p) = p_\alpha p_\beta p_\varsigma p_\delta (p^2)^{-1/2}\frac{1}{512}$$

(7.5)

This expression can be collapsed with metric components to obtain information about the effective theory. For example if we take momentum of the form $p = (p_+, p_-, 0)$ we have

$$U_{++++}(p) = \frac{1}{512}\frac{p_+^3}{p_-}\sqrt{p_+p_-} \text{ and } U_{+-+-}(p) = \frac{1}{512}p_+p_-\sqrt{p_+p_-}$$

(7.6)

Note that the behavior of $U(p)$ in 3d gravity is very different from the two dimensional case studied in [49] because of the square root of $p_+p_-$. To illustrate the difference consider the 2d gravity coupled to a dirac fermion where Alvarez-Gaume and Witten obtained:

$$U_{abcd}(p) = \frac{1}{2} p_a p_b p_c p_d \int_0^1 dx (1-2x)^2 (1-x) x \int \frac{d^2\ell}{(2\pi)^2} \frac{1}{(\ell^2 + x(1-x)p^2)^2}$$

$$= \frac{1}{2} p_a p_b p_c p_d (p^2)^{-1} \frac{1}{(4\pi)} \int_0^1 dx (1-2x)^2 = \frac{1}{24\pi} p_a p_b p_c p_d (p^2)^{-1}$$

(7.7)

So that $U^{(2d)}_{++++}(p) = \frac{1}{24\pi} \frac{p_+^3}{p_-}$ and $U^{(2d)}_{+-+-}(p) = \frac{1}{24\pi} p_+ p_-$. For example collapsing with metric components $\phi = h_{+-}$ one obtains a local action for the 2d Liouville field proportional to $\partial \phi \partial \phi$ as was shown in [49]. Because of the difference between (7.5) and (7.7) one must work harder to study the origin of 2d Liouville theory in 3d gravity plus matter.

If one studies the behavior of 3d gravity interacting with fermionic matter on $M = R^2 \times S^1$ it is possible to study the delicate relation between 3d gravity and 2d gravity with regards to the origin of the 2d Liouville action. In that case we have an infinite sum over of 2d gravity terms. So that (7.7.) is replaced by:

$$U^{R^2 \times S^1}_{abcd}(p) = \frac{1}{2} p_a p_b p_c p_d \int_0^1 dx (1-2x)^2 (1-x) x \sum_{n_3=-\infty}^{\infty} \int \frac{d^2\ell}{(2\pi)^2} \frac{1}{(\ell^2 + (\frac{2\pi n_3}{L_3})^2 + x(1-x)p^2)^2}$$

$$= \frac{1}{8\pi} p_a p_b p_c p_d \int_0^1 dx (1-2x)^2 (1-x) x \sum_{n_3=-\infty}^{\infty} ((\frac{2\pi n_3}{L_3})^2 + x(1-x)p^2)^{-1}$$

(7.8)

Now performing the summation over the discrete momentum $n_3$ we have:

$$U^{R^2 \times S^1}_{abcd}(p) = \frac{1}{24\pi} p_a p_b p_c p_d (p^2)^{-1}$$

$$+ \frac{1}{16\pi} p_a p_b p_c p_d \int_0^1 dx (1-2x)^2 (1-x) x L_3^2 (x(1-x)p^2 L_3^2)^{-1/2} (\frac{\sinh(\sqrt{x(1-x)p^2 L_3^2})}{\cosh(\sqrt{x(1-x)p^2 L_3^2}) - 1} - \frac{2}{\sqrt{x(1-x)p^2 L_3^2}})$$

(7.9)

In the limit of small $L_3$ we have:

$$U^{R^2 \times S^1}_{abcd}(p) = \frac{1}{24\pi} p_a p_b p_c p_d (p^2)^{-1} + \frac{1}{24\pi} \frac{1}{120} p_a p_b p_c p_d L_3^2 + \ldots$$

(7.10)

while for large $L_3$ we recover (7.6) after scaling the gravitational coupling by the radius of $S^1$. Now in this case we have $U^{R^2 \times S^1}_{+-+-}(p) = \frac{1}{24\pi} p_+ p_-$ to leading order so the effective 2d Liouville action of the form $\partial \phi \partial \phi$ is obtained. Again the main point is that one obtains the usual effective action for two 2d gravity from the 3d gravity theory with matter and this useful in showing the between M-theory and type IIA string theory in the 3d gravity approach to M-theory. As conformal dimensions of fields can be defined by the effective coefficient of the induced Liouville term this method allows one to define a critical dimension of M-theory. For example in the conformal gauge for 3d gravity defined by (2.4) the three-metric can be put in the form:

$$g_{\mu\nu} = \begin{pmatrix} e^{\phi_1}\hat{g}_{11} & e^{\phi_1}\hat{g}_{12} & \phi_2 \\ e^{\phi_1}\hat{g}_{21} & e^{\phi_1}\hat{g}_{22} & \phi_3 \\ \phi_2 & \phi_3 & 1+e^{-\phi_1}\hat{g}^{ab}(\phi_a\phi_b) \end{pmatrix}$$

(7.11)

In this form the metric has the symmetry $\phi_1 \to \phi_1 + \alpha \quad \hat{g} \to e^{-\alpha}\hat{g}$. As the 3d gravity reduces to 2d gravity in the $L_3 \to 0$ the implications of this symmetry reduce to the same consistency condition discussed in [2,3]. In particular is one starts with a 3d gravity with $D = 3+N$ target space dimensions and 3d field content:

$$(3,0,g_{\mu\nu}) \oplus (N,N,X^M) \oplus (3,0,\chi_\mu) \oplus (N,N,\psi^M)$$

(7.12)

This reduces in the $L_3 \to 0$ limit to the 2d gravity field content:

$$(1,-1,g_{ab}) \oplus (1,0,a_b) \oplus (1,1,X^9) \oplus (N,N,X^M) \oplus (1,-1,\chi_a) \oplus (1,1,\psi^9) \oplus (N,N,\psi^M)$$

(7.13)

The consistency condition from (7.11) is well known for this system and leads to the equation of total zero conformal dimension from gravity and matter [17]:

$$1+1+N+\frac{1+1+N}{2}+11-26=0$$

(7.14)

So that the critical number of matter fields $N = 8$ determines the critical dimension for the 3d gravity description of M-theory to be $D = 3+N = 11$. The value of $N = 0$ leads to the $c = 1$ 2d gravity consistency condition as discussed in [41] $3Q^2 +1+1+\frac{1}{2}+\frac{1}{2}+11-26=0$ where $Q = 2$. In this case the 3d gravity description of the M-theory that reduces to the type 0A string in the $L_3 \to 0$ limit has $D = 3+N = 3$ target space dimensions.

# VIII Conclusions

In this paper we investigated a 3d gravity approach to M-theory. We have studied the metric components left over after gauge fixing and interpreted these as Liouville modes adding three dimensions to Target space for the theory. We studied the vacuum amplitude on a three torus, $SL(3,Z)$ modular invariance and sum over spin structures to relate the spectrum of eleven dimensional supergravity and type IIA string theory from the 3d gravity approach. We studied the vacuum amplitude on a three annulus to include gauge fields in the 3d gravity M-theory method, which is closely analogous to the inclusion of gauge fields in open string theory. For the case of M-theory in three Target space we use a 3d gravity description to relate it to *type 0A* string theory, or on manifolds with boundary to a $E_8 \times SO(8)$ 2d heterotic string. We used the nonperturbative Chern-Simon description of 3d gravity to go beyond quadratic order and discussed the use of auxiliary fields to build a expansion about $(e,\omega) = 0$ with matter. From the target space point of view, these auxiliary fields can be related to target space momentum and thus could in principle lead to strong coupling phases in certain regimes of momentum space. Finally we studied to the interaction of fermionic matter with 3d gravity at in order to investigate the origins of conformal dimension and Liouville effective action from the 3d gravity perspective.

How does the 3d gravity approach to M-theory relate to other approaches? We have already mentioned the existence of Lattice and Matrix models in the description of nonperturbative pure 3d gravity so that discrete descriptions of M-theory could in principle be related to these. Also the membrane approach to M-theory a three dimensional reparametrisational invariance of the membrane action is related to the conformal invariance of the Nambu-Goto action for a string [51] and this is analogous to the reduction of the 3d gravity description to a 2d Liouville gravity description. The light cone gauge in membrane theory leads to 3+8 split in dynamical fields analogous to the 3+8 split between Liouville and matter fields in the 3d gravity approach. The relation between the background type IIA string Nambu-Goto action and the background eleven dimensional membrane world volume [52] action is analogous to the relation of the spectrum of the type IIA string and eleven dimensional gravity we found by studying the torus amplitude of 3d gravity. Still the presence of the curvature term in the 3d gravity approach means that it will be more difficult to integrate out the metric field to obtain a Nambu-Goto type membrane action than it is for the string or particle where the curvature term is nondynamical. Also, as in the 2d gravity approach to string theory, the 3d gravity approach to M-theory allows the theory to be defined dimensions other than eleven. In three dimensions, in particular, M-theory can be described by the finite and stable world volume Chern-Simons gravity. Finally the 3d gravity approach to M-theory brings together many seemingly separate theories of quantum gravity including gauge gravity, loop space and string models as it exists in a dimension where each of these methods can be fruitfully applied.